\documentstyle[seceq,twocolumn,epsfig]{jpsj}
\def\e{{\rm e}}
\def\virg{\;\;,}
\def\point{\;\,.}

\def\gsim{\,\lower0.75ex\hbox{$\ggs$}\,}
\def\lsim{\,\lower0.75ex\hbox{$\lls$}\,}

\def\Dc{{\cal D}^{\rm crit.}}

\def\Krc{K_{\rho}^{\rm crit.}}
\def \D {{\cal D}}
\def\runtitle{Role of interchain hopping 
 in two  disordered chains of spinless fermions }

\def \runauthor{ Edmond {\sc Orignac}, Yoshikazu {\sc Suzumura} and
Thierry {\sc Giamarchi}}
\title
{
Role of interchain hopping 
 in two  disordered chains of spinless fermions 
 }

\author{
Edmond {\sc Orignac${}^1$ }, Yoshikazu  {\sc Suzumura$^{2,3}$} and Thierry {\sc Giamarchi$^4$}
}   
\inst{
${}^1$ Laboratoire de Physique Th\'eorique, CNRS UMR 8549 Ecole Normale 
  Sup\'erieure, 24 Rue Lhomond 75005 Paris, France \\
${}^2$ Department of Physics, Nagoya University, Nagoya 464-8602,
Japan \\
${}^3$ CREST, Japan Science and Technology Corporation (JST) \\
${}^4$ Laboratoire de Physique des Solides CNRS UMR 8502, Universit{\'e} Paris--Sud,
                  B{\^a}timent 510, 91405 Orsay, France\\
}

\recdate{\hspace{3.5cm} }

\abst
{
 The effect of interchain hopping  on localization 
 has  been investigated for two coupled chains of 
 spinless fermions with  nearest neighbor 
 interaction. 
By use of the renormalization group method, 
 it is found that   interchain hopping plays a significant role  
   when the correlation gap induced by  interplay of  interchain  
   hopping  and intrachain interaction becomes larger 
 than the Anderson localization energy scale induced by impurities.
 The strong competition between  localization and  delocalization 
 is examined for  attractive interaction. 
 }

\kword
{
spinless fermion, interchain hopping, 
Anderson localization, metal-insulator transition, two coupled chain
}
\begin{document} 
\maketitle


%
\section{Introduction}
\begin{figure}
 \begin{center}
\epsfig{file=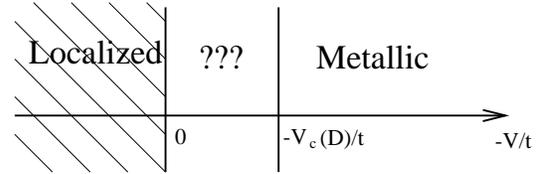,angle=0,width=7cm}
 \end{center}
\caption{The phase diagram derived in
{Ref.~\citen{orignac_2chain_long}} For repulsive interactions, a
localized insulating 
phase is obtained. For attractive interactions, a metallic phase is
obtained. In the region $V\in [V_c,0]$ the localization length is of
the order of magnitude of the correlation length of the gapped
excitations of the system. This transition region cannot be described
with the methods\cite{orignac_2chain_long} used to describe the two
other regions}
\label{fig:previous}  
\end{figure}

The interplay of disorder and interactions in low dimensionality 
is one of the most
important topics in condensed matter physics. In the recent years, it
has known a renewed interest due to experimental evidences for a metal
insulator transition in a two dimensional electron gas
\cite{kravchenko_mit_2d} in the absence of a magnetic field. Such a
transition being ruled out for non-interacting
electrons\cite{abrahams_loc}, the interplay of disorder and
interaction is believed to be responsible for the transition. However,
the
theoretical study of the interplay of disorder and interaction in two
dimensions is an extremely difficult
topic\cite{belitz_localization_review} and this problem remains
extremely controversial. There exists also quasi-one dimensional
systems in which the interplay of disorder and interactions can be
studied experimentally, such as quantum wires\cite{tarucha_quant_cond}
and nanotubes\cite{nano1}. 
 In one dimension, the theoretical
situation is better since bosonization
techniques\cite{solyom_revue_1d} can be used to deal with
interactions, allowing for a Renormalization Group (RG) treatment of
disorder\cite{giamarchi_loc}. 
It has been found that for a one-chain
system, attractive interactions of the order of the bandwidth were
needed to suppress Anderson localization and produce a metallic
state\cite{giamarchi_loc}.    
More recently, it was found that in a model of two coupled spinless
fermions chain\cite{orignac_2chain_short,orignac_2chain_long} the
opening of gaps in some excitations of the system could lead to a
drastic modification of the response of the system to disorder. 
Namely, in the case when disorder is too small to close the gaps of
the spinless fermion ladder, the system is delocalized for attractive
interactions (see figure \ref{fig:previous}). 
As a result, for small enough disorder, the metal
insulator transition occurs for weakly attractive interactions. 
However,  the
approximation of disorder weak enough not close the gaps of the system
 used in
Refs. \citen{orignac_2chain_short,orignac_2chain_long} is not valid
 close to the non-interacting point where the gaps vanish, precluding 
 a quantitative description of the metal insulator
transition in this regime within this approximation.
Fortunately, the fact that the transition occurs for weak disorder and
weak interactions implies
that the techniques of perturbative renormalization group can be
applied to describe the transition. In the present paper, we undertake
such an RG analysis. We will show that there exist three regimes. For
sufficiently repulsive interactions and sufficiently large interchain
hopping, there is a regime of pinned $CDW^\pi$ in which interchain
hopping is relevant\cite{orignac_2chain_long}. For attraction strong
enough, there is a metallic phase in which $t_\perp$ is
relevant\cite{orignac_2chain_long}. Finally, for weak interactions or
small interchain hopping, there is a regime in which the system
behaves as a single chain insulator and interchain hopping becomes an
irrelevant perturbation.

The plan of the paper is as follows: 
In section \ref{sec:model_rge} 
we recall the definition of the model. Then, we set up the bosonized
Hamiltonian and the renormalization group equations. In section
\ref{sec:results} we
discuss the physics obtained from solving the RG equations. We
identify the fixed points of the RG equations and discuss the nature
of the corresponding phases of the two-chain system. Then, we determine the
phase diagram resulting from the RG equations. Finally, we discuss the
behavior of the localization length with disorder strength. 
In section \ref{sec:conclusion}, we summarize our results and discuss
possible extensions. Technical details have been left to Appendixes
\ref{app:ope_approach} and \ref{app:criterion}.

\section{Model and Renormalization Group
  Equations}\label{sec:model_rge} 
\subsection{Hamiltonian}
 We consider a system of two coupled 
 chains of spinless fermions, described by
the Hamiltonian:
\begin{eqnarray}
                      \label{Hamiltonian}
 {\cal H} & = & 
 -t \sum_{i,l} \left( c^{\dagger}_{i,l} c_{i+1,l} + {\rm h.c.} \right)
    - t_{\perp} \sum_{i}\left(c^{\dagger}_{i,1}c_{i,2}+
      c^{\dagger}_{i,2}c_{i,1} \right)  \nonumber \\
  & &     + V \sum_{i} n_{i,l} n_{i+1,l}
       -   \sum_{i}\left( \xi_{i,1} c^{\dagger}_{i,1}c_{i,1}+
       \xi_{i,2} c^{\dagger}_{i,2}c_{i,2} \right) \virg \nonumber \\
\end{eqnarray}
 where $c_{i,n}$ ($n$=1,2) denotes a fermion operator in the  $n$-th chain.
 The lattice constant is taken as unity. 
Quantities   $t$ and $t_{\perp}$ denote  intrachain hopping energy and 
  interchain hopping energy respectively ( $t > t_{\perp}$)
 and $V$ is the interaction  between nearest neighbor sites.  
 Quantities $\xi_{i,1}$ and  $\xi_{i,2}$ are the random potential 
 acting on the chain 1 and the chain 2 respectively. One has:
$\overline{\xi_{i,p} \xi_{i',p'}}=W \delta_{i,i'}\delta_{p,p'}$. 
In the following, we assume that $t_\perp$ is small enough that there
are two bands at the Fermi level. If $\rho$ is the average number of
fermions per site, this means that $t_\perp \ll 2 t \sin^2 (\pi
\rho)$. We assume incommensurate filling.

By diagonalizing the $t_{\perp}$-term  and making use of 
 the  bosonization, 
 eq.(\ref{Hamiltonian})  is rewritten as 
\cite{nersesyan_2ch,orignac_2chain_long}
\begin{eqnarray}
             \label{PH}
 H &= & H_\rho + H_\parallel + H_{\rm{dis.}}   \virg  \\
  H_\rho & =& \int \frac{dx}{2\pi} \left[ u_\rho K_\rho (\pi \Pi_\rho)^2 
        +\frac{u_\rho}{K_\rho} (\partial_x \phi_\rho)^2 \right] 
                     \virg  \\
 H_\parallel & =& \int \frac{dx}{2\pi} \left[ u_\parallel K_\parallel (\pi \Pi_\parallel)^2  
   +  \frac{u_\parallel}{K_\parallel} (\partial_x \phi_\parallel)^2 \right]
   \nonumber \\ 
 & + & \frac{2g_f}{(2\pi \alpha)^2} \int dx \cos \sqrt{8} \theta_\parallel \nonumber \\ 
     & + &  \frac{2g_\perp}{(2\pi \alpha)^2} \int dx \cos (\sqrt{8}
               \phi_\parallel + mx) \virg  \\
 H_{\rm{dis.}} & = & \int dx \left[ \frac{\xi_s(x)}
         {\pi a} e^{i \sqrt{2}
        \phi_\rho} \cos(\sqrt{2}\phi_\parallel +\frac {mx} 2)
   \right.  \nonumber  \\ 
  & & \left.  +\frac{\xi_a(x)}{\pi a} e^{i \sqrt{2}
   \phi_\rho} \cos \sqrt{2}\theta_\parallel + \rm{H. c.} \right] 
                  \virg
\end{eqnarray}

 In Eq.(\ref{PH}), random potentials expressed  in terms of 
   the symmetric and antisymmetric parts 
    are treated as  
     $\overline{\xi_a(x)\xi_a^*(x')}=D_a \delta(x-x')$ and
$\overline{\xi_s(x)\xi_s^*(x')}=D_s \delta(x-x')$. We have neglected
forward scattering since it does not contribute to Anderson
localization \cite{abrikosov_rhyzkin}. 
   The quantity  $m=-\frac{4K_\parallel t_\perp}{u}$ measures 
  the difference of Fermi wavevectors between
           the bonding and the antibonding band. The parameters
$K_\rho,K_\parallel, u_\rho, u_\parallel, g_f,g_\perp,D_s,D_a$ are related to
$t,t_\perp,V,W$ by: 
\begin{eqnarray}
\label{K_ug}
K_\parallel=1,\nonumber \\
u_\parallel=v_F(1+\frac{Va}{\pi v_F}(1-\cos(2k_F
a))), \nonumber  \\
g_f=-Va(1-\cos(2k_F a)),\nonumber\\
g_\perp=- Va(1-\cos(2k_F a)),\nonumber\\
u_\rho = v_F(1+\frac{Va}{\pi v_F}(1-\cos(2k_F a))),
\nonumber  \\
K_\rho = 1 -\frac{Va}{\pi v_F}(1-\cos(2k_F a)),\nonumber\\
D_s=D_a=\frac{Wa}{2},
\label{param}
\end{eqnarray} 
where $a$ is a lattice spacing and $v_F=2ta \sin(k_Fa)$ is the Fermi
velocity of a single chain\cite{orignac_2chain_long}.
The bosonized Hamiltonian in the absence of disorder has been
previously derived by Nersesyan et al.~\cite{nersesyan_2ch} by mapping the
problem onto a Luttinger liquid with interactions breaking spin
rotation symmetry in a magnetic field~\cite{giamarchi_spin_flop}. 

\subsection{Renormalization Group Equations}
It will be convenient to write the RG equations in terms of the following
  dimensionless quantities: 
  \begin{eqnarray}\label{eq:dimensionless}
  y_f=\frac{g_f}{\pi u_\parallel} ,\; 
       y_\perp=\frac{g_\perp}{\pi u_\parallel} ,
        \; \tilde{t}_\perp=\frac{t_\perp a}{u_\parallel} 
                              \virg \nonumber \\
{\cal D}_a=\frac{D_a a}{\pi u_\rho^2}, \; 
  {\cal D}_s=\frac{D_s a}{\pi u_\rho^2} \point
\end{eqnarray}
In the following, we will take $u_\rho=u_\parallel=v_F$. 
 By taking into account the effect of  interchain hopping, 
  the RG equations are derived as: (See Appendix \ref{app:ope_approach})
\begin{subeqnarray}\label{eq:RGE_full}
\frac{d K_\rho}{dl}& =-&K_\rho^2 ({\cal D}_a + {\cal D}_s)
                           \virg  \\
\frac{d K_\parallel}{dl} & = & {\cal D}_a -K_\parallel^2 {\cal D}_s
+ \frac 1 2 (y_f^2 - K_\parallel^2 y_\perp^2 J_0(4K_\parallel\tilde{t}_\perp)) 
                   \virg \nonumber  \\ \\
\frac{dy_f}{dl} & = &\left(2-\frac 2 {K_\parallel}\right) y_f -2 {\cal
D}_a  
                         \virg  \\
\frac{dy_\perp}{dl} & = &\left(2- 2 K_\parallel\right) y_\perp -2 {\cal
D}_s 
                           \virg  \\
\frac{d{\cal D}_s}{dl} & = & \left(3-K_\rho-K_\parallel-y_\perp
\right){\cal D}_s   
                         \virg  \\ 
\frac{d{\cal D}_a} {dl} & = & \left(3-K_\rho-\frac 1
{K_\parallel}-y_f \right){\cal D}_a  
                            \virg \\ 
\frac {d\tilde{t}_\perp}{dl} & =& \tilde{t}_\perp - \frac{y_\perp^2} 4
J_1(4K_\parallel\tilde{t}_\perp) \virg 
\end{subeqnarray}
 which  are calculated with the initial conditions,
$y_f(0) = y_{\perp}(0) = K_{\rho} - 1 = -  \frac{V}{\pi t} \sin(k_F a)$. 
For ${\cal D}_a={\cal D}_s=0$, these RG equations are identical to the
ones of the pure system
\cite{nersesyan_2ch,giamarchi_spin_flop,ledermann_2band_spinless}.

\subsection{The case $\tilde{t}_\perp = 0$}
 In the limit of  $\tilde{t}_{\perp}=0$, a simplified form of 
  the  RG equations (\ref{eq:RGE_full}) can be obtained. With initial
conditions  $K_{\parallel}(0) = 1$, $y_f(0)=y_{\perp}(0)$,
$\D_a(0)=\D_s(0)$ and $\tilde{t}_\perp=0$, it is easily seen that one
has $K_\parallel(l)=K_\parallel(0)=1$, $y_f(l)=y_\perp(l)$ and
$\D_s(l)=D_a(l)$, $t_\perp(l)=0$ for any $l$.  
By using  $y_f =y_\perp =  K_{\rho} - 1$, it is seen that the RG
equations (\ref{eq:RGE_full}) can be reduced to a pair of RG equations
for $y_f$ and $\D=\D_{a,s}$.   
 the single chain RG equations are  written as
\begin{eqnarray}\label{eq:RGE_single_chain}
 \frac{d y_f}{d l} &=& - 2  \D 
         \virg  \\
 \frac{d \D}{dl} &=& (1-2 y_f) \D   \point
\end{eqnarray} 
These RG equations are identical to the RG equations for the
single spinless fermions chain\cite{giamarchi_loc}:
\begin{eqnarray}
 \frac{d K_\rho}{d l} &=& - 2 K_\rho^2  \D 
         \virg  \\
 \frac{d \D}{dl} &=& (3-2K_\rho) \D  \point
\end{eqnarray} 
in the weak interaction limit where $K_\rho\simeq 1+y_f$.
 As a result, for $t_\perp=0$ we recover the
RG equations of a single spinless fermion chain in the vicinity of the
non-interacting point. Let us
note that the scaling approach will not reproduce the prefactor in the
localization length that is obtained in the case of non-interacting
 electrons\cite{prigodin_firsov}.    
From the limit $t_\perp=0$, we expect that for $\tilde{t}_\perp$ small
 enough, the two chain system will behave identically to a single
 chain system. For larger $t_\perp$, a crossover to the ladder
 regime~\cite{orignac_2chain_long} should occur. 
\subsection{The non-interacting limit}
As in the case of the single chain, the non-interacting limit has to
be analyzed very carefully. It is known that in the single chain case, 
disorder seems to induce an effective interaction in the
system\cite{giamarchi_loc} even in the absence of interaction in the
pure system. This artefact can be cured
by separating properly disorder and inelastic processes\cite{giamarchi_loc} , so that starting from a
non-interacting problem, one remains with a non-interacting problem. 
The effective interaction can be evaluated in the case of the two
chain problem. If we define as $K_\rho^{({\rm dis.})},
K_\parallel^{({\rm dis.})}, y_f^{({\rm dis.})}, y_\perp^{({\rm
    dis.})}$ the corrected initial values of
$K_\rho,K_\parallel,y_f,y_\perp$ in the presence of disorder, we have:
\begin{eqnarray}
  \label{eq:initial_conditions_proper}
  K_\rho^{({\rm dis.})}=K_\rho -\D_a -\D_s \\
  K_\parallel^{({\rm dis.})}=K_\parallel +\D_a -\D_s \\
  y_f^{({\rm dis.})}=y_f - 2 \D_a \\
  y_\perp^{({\rm dis.})}=y_\perp -2\D_s
\end{eqnarray}

With these initial values, it can be shown easily that if we start
from $\D_a=\D_s$ and a non-interacting system, we  will preserve
$K_\parallel=1, K_\rho=1, y_f=y_\perp=0$ under the RG flow and thus no
spurious interaction is generated.

\section{Competition of disorder and interchain hopping}\label{sec:results} 

\subsection{Analysis of the RG flow}
In the present section, we discuss the strong coupling fixed points of
the renormalization group equations (\ref{eq:RGE_full}). We show that
there are three types of strong coupling fixed points. The first one
is associated with the pinned $CDW^\pi$ of the two-chain
system\cite{orignac_2chain_long}, the second one is associated with
Anderson localization in the single chain\cite{giamarchi_loc} i. e. to
an irrelevant $t_\perp$ and the last one is associated with the metallic
phase of the two chain system\cite{orignac_2chain_long}. In the
following subsections, we discuss in more details the different fixed
points. In section \ref{sec:transition_qualitative}, we discuss
qualitatively the transitions between these three phases. 
 
\subsubsection{$\D(l)\to \infty,y_f(l)\to -\infty, K_\parallel(l) \to
  + \infty$} 
\begin{figure}
\begin{center}
{\epsfig{file=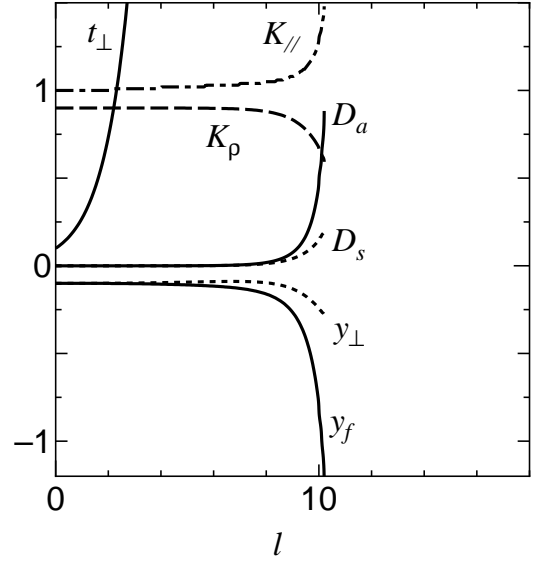,angle=0,width=7cm}}
\end{center}
\caption{
 The $l$-dependence of 
 $t_{\perp}(l)$, $K_{\rho}(l)$, $K_{\parallel}(l)$, 
 $y_{\perp}(l) $, $y_f(l)$, $\D_s(l)$ and $\D_a(l)$
 for $t_{\perp}(= t_{\perp}(0) ) =0.1, \D ( =\D_s(0) ) =\D_a(0)=10^{-6}$ 
and $K_{\rho} ( =K_{\rho}(l) ) =0.9$.
 }
\label{fig1a}
\end{figure}

This regime is represented on figure \ref{fig1a}.
 We have $K_\parallel \to +\infty$ so that
$\theta_\parallel$ becomes classical. Since $y_f \to -\infty$, the
expectation value of $\theta_\parallel$ is $\langle \theta_\parallel
\rangle =0$. This corresponds to the formation of the $CDW^\pi$ in the
pure system, thus this regime corresponds to the pinned $CDW^\pi$
obtained for repulsive interactions\cite{orignac_2chain_long}. 
We define the correlation length associated with this regime by: 
  $1/\xi_\parallel \equiv   {\rm e}^{- l_\parallel}$ (correlation
  length of $\theta_\parallel$) and  
   $1/\xi_{\rm loc.}  \equiv  {\rm e}^{- l_{imp}}$ (localization length) 
 where 
 $y_f(l_\parallel) = 1$ and $\D_a(l_{imp}) = 1$.   
  The energy scales  corresponding to these correlation lengths 
    are given by  
     $\Delta_\parallel = v_f {\rm e}^{- l_\parallel}$ and  
    $E_{{\rm loc.}} = v_f {\rm e}^{- l_{imp}}$,  respectively. 
Within this regime, there are two possibilities.

 The first one, shown
on Fig. \ref{fig1a} is that
$y_f$ and $\D_a$ reach strong coupling simultaneously. This regime can
be identified with the single chain regime by noting that
Eq. (\ref{eq:RGE_single_chain}) that describes the regime $t_\perp=0$ 
 has the same type of strong coupling
fixed point. Let us note that a regime where $\D_a$
reaches strong coupling before $y_f$ is excluded since in
Eqs. (\ref{eq:RGE_full}) $\D_a$ contributes to the renormalization of
$y_f$.

\begin{figure}
\begin{center}
{\epsfig{file=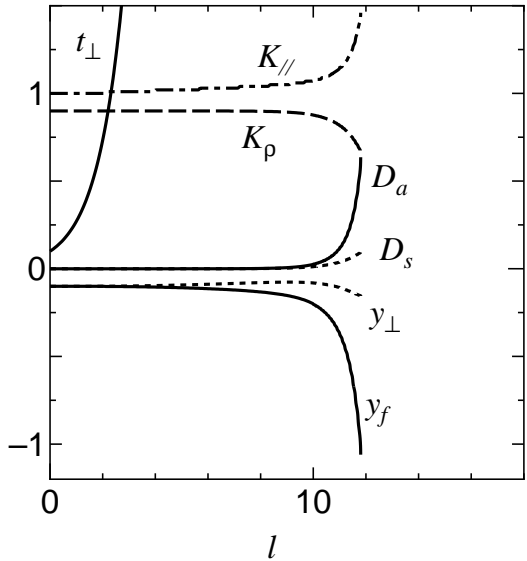,angle=0,width=7cm}}
\end{center}
\caption{
 The $l$-dependence of
 $t_{\perp}(l)$, $K_{\rho}(l)$, $K_{\parallel}(l)$, 
 $y_{\perp}(l) $, $y_f(l)$, $D_s(l)$ and $D_a(l)$
 for $t_{\perp}=0.1, D=10^{-7}$ 
and $K_{\rho} =0.9$.
 }
\label{fig1b}
\end{figure}

The second possibility, shown on Fig. \ref{fig1b}, corresponds to $y_f$
reaching strong coupling before $\D_a$. In this regime, to obtain the
localization length, one needs to renormalize up to the scale $l^*$
where $y_f(l^*)=1$, then use the simplified renormalization group
equations\cite{orignac_2chain_long} that are valid when the
expectation value of $\theta_\parallel$ is well defined. This regime
corresponds obviously to the pinned $CDW^\pi$
regime\cite{orignac_2chain_long}.     

\subsubsection{$y_f(l) \to +\infty, K_\parallel(l) \to +\infty, \D(l)
  \to 0$}

\begin{figure}
\begin{center}
{\epsfig{file=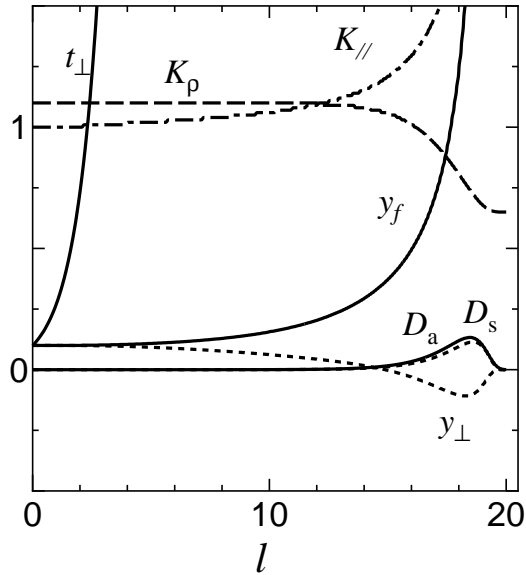,angle=0,width=7cm}}
\end{center}
\caption{
 The $l$-dependence of
 $t_{\perp}(l)$, $K_{\rho}(l)$, $K_{\parallel}(l)$, 
 $y_{\perp}(l) $, $y_f(l)$, $D_s(l)$ and $D_a(l)$
 for $t_{\perp}=0.1, D=10^{-7}$ 
and $K_{\rho} =1.1$.
 }
\label{fig1c}
\end{figure}
This regime is represented on Fig. \ref{fig1c}.
The field $\theta_\parallel$ becomes  classical,
but now its expectation value in the ground state of the system is
$\frac{\pi}{\sqrt{8}}$, making disorder irrelevant. 
In the pure system, this corresponds 
to the s-wave superconductor\cite{orignac_2chain_long} ($SC^s$)
obtained for attractive interactions and
there should be \emph{no} Anderson localization. This is indeed what
is obtained since disorder is scaled to zero(cf. Fig.\ref{fig1c}). 
We define $K_\rho^{\rm
  crit.}(t_\perp,\D)$ as the value of $K_\rho$ above which this regime
is obtained. 

\subsection{Transition between the different
  regimes}\label{sec:transition_qualitative}

We  discuss qualitatively the crossovers between the different phases
 in the system. 

\begin{figure}
\begin{center}
{\epsfig{file=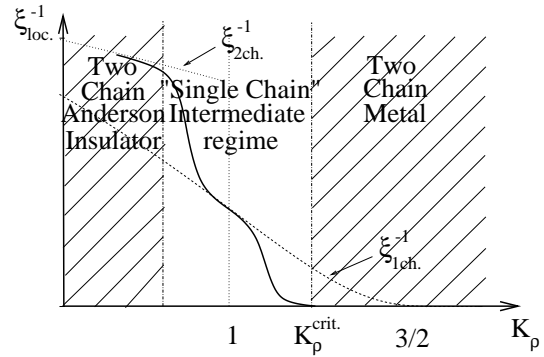,angle=0,width=7cm}}
\end{center}
\caption{The behavior of the localization length $\xi_{{\rm loc.}}$
as a function of $K_\rho$ for fixed $t_\perp$ and $D$. For $K_\rho$
small enough, the gap in $\theta_\parallel$ is well formed and we have
the pinned $CDW^\pi$ or two chain Anderson
Insulator\cite{orignac_2chain_long}. For $K_\rho$ large enough, the
gap in $\theta_\parallel$ is also robust to disorder and we obtain the
two-chain metal\cite{orignac_2chain_long}. In the intermediate regime,
disorder can compete with the formation of a gap in $\theta_\parallel$
and establish a single chain regime. }
\label{fig:loclengths}
\end{figure}

We have to compare three correlation lengths (or the associated energy
scales), the single chain localization length $\xi_{{\rm 1 ch.}}$, the two chain
localization length $\xi_{{\rm 2 ch.}}^{V>0}$ and the correlation length in the antisymmetric
modes $\xi_\parallel$. 
It is clear that the two-chain localization length in the repulsive
regime $\xi_{{\rm 2 ch.}}^{V>0}$ is always shorter than the single
chain length $\xi_{{\rm 1 ch.}}$. Let us first consider the repulsive
regime. 
  For $V,t_\perp$ large enough and $D$ weak
enough, we expect $\xi_\parallel$ to be the shortest length in the
problem.  In this regime, the calculations of
Ref. \citen{orignac_2chain_short,orignac_2chain_long} are valid
and the localization length in the system is $\xi\sim \xi_{{\rm 2
ch.}}^{V>0}$. This is the regime represented on Fig.~\ref{fig1b}.  
Let us now imagine that we are decreasing $t_\perp$ or $V$ so as to
increase $\xi_\parallel$. We will first reach a regime where
$\xi_\parallel \sim \xi_{{\rm 2
ch.}}^{V>0}$. In this regime, the calculations of
Ref. \citen{orignac_2chain_short,orignac_2chain_long} are not
valid anymore and the localization length $\xi$ starts to cross over to the
single chain regime, i.e. $\xi_{{\rm 2
ch.}}^{V>0}<\xi <\xi_{{\rm 1 ch.}}$. This case corresponds to figure
\ref{fig1a}. Finally, as $\xi_\parallel \sim
\xi_{{\rm 1 ch.}}$, the correlation length falls into the single
chain regime $\xi=\xi_{{\rm 1 ch.}}$. This is in particular the
regime that obtains for $t_\perp=0$ or $V=0$. Suppose now that we
start from $V=0$ and decrease $V$. Then, we are in the attractive
regime, and $\xi_\parallel$ starts to decrease. When
$\xi_\parallel\sim \xi_{{\rm 1 ch.}}$, we fall into the two chain
regime again. But this time there is no Anderson localization at all,
and $\xi=\infty$. This is the regime of Fig.~\ref{fig1c}. 
These successive transitions for fixed $t_\perp$ are
conveniently described as a function of $K_\rho$ on figure
\ref{fig:loclengths}.  

\begin{figure}
\begin{center}
{\epsfig{file=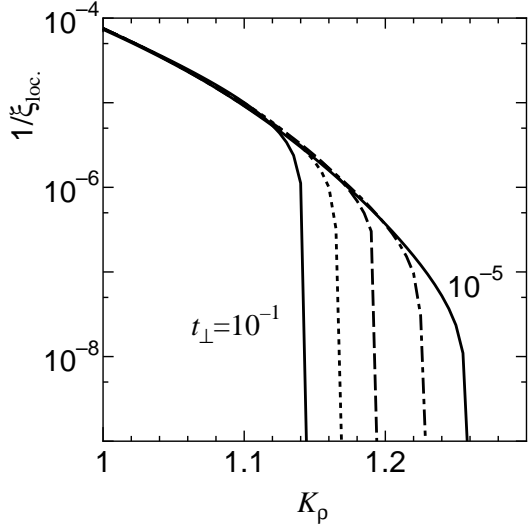,angle=0,width=7cm}}
\end{center}
\caption{
The $K_{\rho}$-dependence of $1/\xi_{\rm loc.}$ 
   and $1/\xi_\parallel$ (dotted curve) with  fixed 
     $t_{\perp} = 10^{-1}$(solid curve)   
        $10^{-2}$(dotted curve), $10^{-3}$(dashed  curve),
           $10^{-4}$(dash-dotted curve) and   $10^{-5}$(solid  curve)
             and fixed $D=10^{-5}$. 
 }
\label{fig2c}
\end{figure}

The crossover from one chain to two chains can be evidenced easily in
the numerical solution of the RG equations. 
In Fig. \ref{fig2c}, 
 $1/\xi_{\rm loc.}$ as a function of $K_{\rho}$ is shown with fixed 
  $D=10^{-5}$ and $t_{\perp} =10^{-1}, \cdots ,10^{-5}$ where   
  only the regime for  $K_\rho>1$ is shown. 
  For $K_\rho-1$ small enough, all curves display the same 
  behavior, which is the one
    obtained for $t_\perp=0$. In this range of $K_\rho$, disorder
 causes Anderson localization at a lengthscale much shorter than the
 correlation length of $\theta_\parallel$ and the system is 
  in the single chain regime. When $K_\rho-1$ is larger, the localization
  length is increased by the combination of $t_\perp$ and the stronger
 attractive  interaction. The change of behavior of
$\xi_{\rm loc.}$ from single chain to ladder occurs earlier when 
  $t_\perp$ is increased. 
 For  $K_\rho > K_\rho^{{\rm crit.}}(t_\perp)$,  one has
  $\xi_{\rm loc.}=\infty$. 
 We remark that (as could be
  expected)  $K_\rho^{{\rm crit.}}$ is a decreasing function of
  $t_\perp$. 

It has been well known that in a spinless fermion system
 the energy scale $E_{{\rm loc.}}=\frac {v_F}{\xi_{{\rm loc.}}}$   
  for  Anderson localization is enhanced  by  repulsive interaction 
 and is suppressed by  attractive interaction. 
  The latter case in the two-coupled spinless fermion chains system
   results in 
   a strong competition between $\Delta_{\parallel}$ and 
  $E_{{\rm loc.}}$. 
 Actually, by comparing  $\Delta_{\parallel}$ in the absence of the impurity 
 with  $E_{{\rm loc.}}$ in the absence of the interchain hopping, 
 we obtain 
  the boundary between the localization and the delocalization  as
(see Appendix \ref{app:criterion} for a derivation)
\begin{equation}
   \label{boundary}
\D = C_0 \bigg( 4 t_\perp \exp \Big[- \frac{\pi}{2 K-2}+1
               \Big] \bigg)^{(3 - 2 K)}
 \virg
\end{equation}
 where $C_0$ is a constant. $K_\rho^{\rm crit.}(t_\perp,\D)$ can be
 obtained  by solving  Eq. (\ref{boundary}) for fixed $t_\perp,\D$. 

If we compare  the transition in the two chain system with
the one in the single chain system, we note in the latter case that the
transition is controlled by a single critical point at
$K_\rho=3/2,\D=0$. As a result, universal exponents are obtained
at the metal insulator transition. This is not anymore the case in the
two chain system, and we expect non-universal exponents at the
metal-insulator transition. These exponents might not be related in a
simple way to $K_\rho^{\rm crit.}$ since $K_\rho^{\rm crit.}$ only
control the exponents associated with charge fluctuations. 
It might be interesting to consider the
dependence of such exponents of the various correlations with $t_\perp$
and $\D$. This will be left for a future study. 

\begin{figure}[htbp]
  \begin{center}
    \epsfig{file=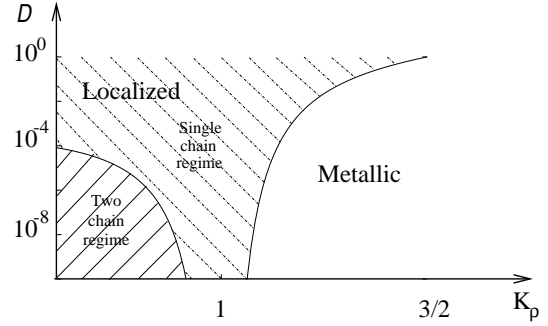,angle=0,width=7cm}
    \caption{The phase diagram predicted from scaling arguments
      Eq. (\ref{boundary}) and Eq. (\ref{boundary_repulsive}). The
      line separating the two localized regime is a crossover line.}
    \label{fig:phase_diagram}
  \end{center}
\end{figure}

Following a similar line of thought to the one that lead to
Eq. (\ref{boundary}), the
boundary between the single chain localization regime and the two
chain localization regime is obtained as:
 \begin{equation}
   \label{boundary_repulsive}
\D = C_1 \bigg( 4 t_\perp \exp \Big[- \frac{\pi}{2-2 K}+1
               \Big] \bigg)^{(3 - 2 K)}
 \virg
\end{equation}
where $C_1$ is another constant. 
The corresponding phase diagram is plotted on figure
\ref{fig:phase_diagram}. 

%

\begin{figure}
\begin{center}
{\epsfig{file=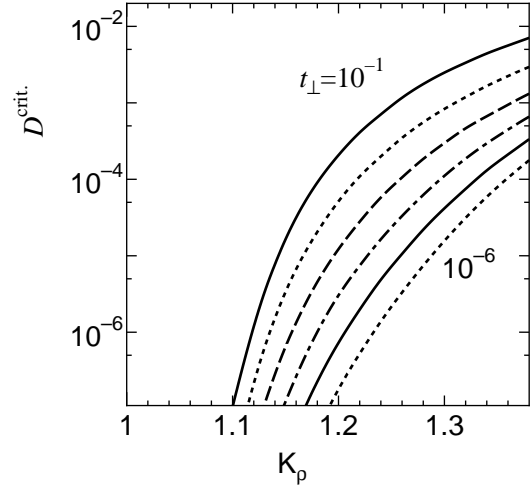,angle=0,width=7cm}}
\end{center}
\caption{  A plot of  $\Dc$ vs. $K_{\rho}$ 
 with the fixed 
 $t_{\perp} = 10^{-1}$ (solid curve), $10^{-2}$ (dotted curve), 
  $10^{-3}$ (dashed curve), $10^{-4}$(dash-dotted curve),
   $10^{-5}$(solid curve) and 
     $10^{-6}$(dotted curve).  
 The dots for $t_{\perp} = 10^{-1}$ and $10^{-6}$ 
   are calculated from eq.(\ref{boundary}). 
 }
\label{fig5}
\end{figure}

We can compare the predictions of these  scaling arguments
with the results of the numerical solution of the RG equations
(\ref{eq:RGE_full}). 
In Fig. \ref{fig5}, such a comparison is shown.  
 The localized region increases  with decreasing $t_{\perp}$.
 These results are  compared with 
   those  obtained by eq.(\ref{boundary}) 
 where $C_0 = 0.0207$ is chosen so as  to fit the data 
 for $t_{\perp}=0.1$. 
  It turns out that 
   this  formula fits well 
   especially for $D < t_{\perp}/10$.

The effect of $t_\perp$ on $K_\rho^{\rm crit.}(\D)$ defined as the
value of $K_\rho$ below which the system is localized for a given
disorder strength $\D$ can also be analyzed using the RG equations.

       \begin{figure}
\begin{center}
{\epsfig{file=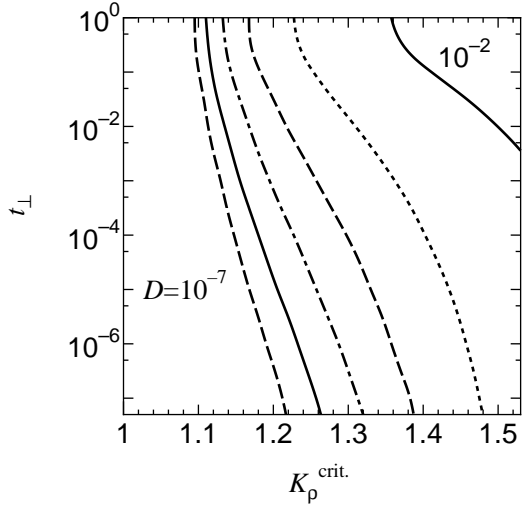,angle=0,width=7cm}}
\end{center}
\caption{
A plot of  $t_{\perp}$ vs. $\Krc$ 
 with the fixed 
 $D = 10^{-2}$ (solid curve), $10^{-3}$ (dotted curve), 
  $10^{-4}$ (dashed curve), $10^{-5}$(dash-dotted curve), 
   $10^{-6}$(solid curve) and  $10^{-7}$(dashed curve). 
 }
\label{fig3}
\end{figure}
      
In Fig. \ref{fig3}, 
 $K_\rho^{{\rm crit.}}$ as a function of $t_\perp$ is shown 
    for various disorder strengths ${\cal D}=10^{-7}\ldots
    10^{-2}$. 
   For  $t_\perp \to 0$, $K_\rho^{{\rm crit.}}$  goes to
     the single chain value $3/2$. As $t_\perp$ is
increased,$K_\rho^{{\rm crit.}}(t_\perp)$    decreases to 
$K_\rho^{\rm crit. *}> 1$. The precise value of this quantity could in
principle be obtained by setting $t_\perp(0)=\infty$ in
Eqs. (\ref{eq:RGE_full}). 
With increasing $D$,  $\Krc$ increases and  
   the regime for localization is enlarged.

\subsection{behavior of the localization length with disorder
 strength} 

\begin{figure}
\begin{center}
{\epsfig{file=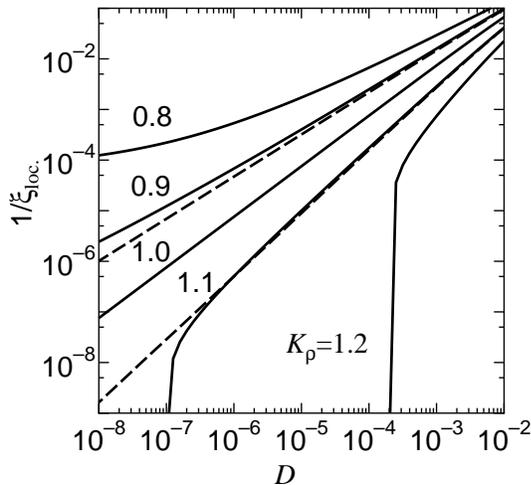,angle=0,width=7cm}}
\end{center}
\caption{
The quantity $1/\xi_{\rm loc.}$ as a function of $\D$ with the fixed 
 $K_{\rho} =$ 0.8, 0.9, 1.0, 1.1 and 1.2 where $t_{\perp}=0.1$.
 For the comparison, $1/\xi$ in the limit of small $t_{\perp}$
  is shown by the dashed curve for $K_{\rho} =$ 0.9 and 1.1.  
 }
\label{fig2a}
\end{figure}

In Fig. \ref{fig2a}, the quantity $1/\xi_{\rm loc.}$ is plotted 
as a function of $\D$ is  
 for fixed values of $K_{\rho}$. 
 For repulsive interactions ($K_{\rho} <1$), 
we see that $1/\xi_{\rm loc.}$ is always enhanced by
 $t_{\perp}$ in qualitative agreement with
 Ref. \citen{orignac_2chain_long}. For $K_{\rho} =1$  corresponding to 
  the absence of intrachain  interaction,   
 the effect of  $t_{\perp}$ on $1/\xi_{\rm loc.}$ is negligible. This is in
 agreement with the fact that for non-interacting fermions the
 localization length is only enhanced by a factor 2 by going from one
 chain to two chains\cite{prigodin_firsov}. 
The solid curve for $K_{\rho} = 1.1$ (attractive interactions) 
 shows a transition from an Anderson localized to a superconducting state 
  at $\D = \Dc \simeq 0.12 \times  10^{-6}$ i. e. between
  Fig. \ref{fig1b} ($\D=10^{-6}$)  and Fig. \ref{fig1c}
($\D=10^{-7}$).   
   Comparing with the single chain case $t_\perp=0,K_\rho=1.1$ (dashed
   curve), we see that  well above  $D \simeq  \Dc$   
   $1/\xi_{\rm loc.}$ is enhanced slightly. Thus, in this regime  
   the interchain hopping enhances  localization even for attractive
   interactions.  This is a consequence of 
a change of  sign of $y_{\perp}$ and $y_f$ that occurs at large
 lengthscales , which implies that at
large enough lengthscales the system behaves as if it were
repulsive. This effect could not be captured by the approximations of
Ref. \citen{orignac_2chain_long} since they only apply to
infinitesimal disorder. 
Finally, we note that for stronger attractive interactions,
$K_\rho=1.2$, the disorder strength needed to cause a transition from
the metal to the Anderson insulator is increased as could be
expected. Note however that this increase is by 3 orders of magnitude
although interaction strength is only doubled. Thus, we see that the
RG allows us to recover the qualitative behavior obtained for the weak
disorder case\cite{orignac_2chain_long}, but also allows us to analyze
in details the interplay of disorder and interaction.

\section{Conclusion}\label{sec:conclusion}

We have presented the results of a renormalization group approach to
the two chain spinless fermion system with disorder. This approach
confirms the results of Ref.~\citen{orignac_2chain_long} in the
limit of disorder small enough not to modify the gaps of the pure
system, namely the strong reinforcement of Anderson localization by
interchain hopping in the case of repulsive interactions and the
delocalization by interchain hopping in the case of attractive
interactions. It also enabled us to obtain new results in the regime
where disorder and interactions are of the same order of magnitude. 
In the latter regime, a strong enough disorder can suppress completely
interchain hopping and maintain the system in a single chain regime. 
In the repulsive regime, disorder, interchain hopping and interaction
reinforce each other leading in particular to an enhancement of the
gap to interchain charge excitations with respect to the pure system. 
In the attractive regime, disorder and interchain hopping compete with
each other. This leads in the delocalized phase to a reduction of the
gap to interchain excitations by disorder. Finally, an exciting effect
is the existence of delocalization by attractive interactions at small
lengthscales followed by Anderson localization at large
lengthscales. This might lead to a non-monotonic behavior of the
charge stiffness of a disordered spinless fermion ladder as a function
of its length. It might prove worthwhile to investigate such an effect
numerically. Another interesting subject for future studies is the
behavior of critical exponents at the localized-delocalized
transition.

\vspace{0.5cm}

{\bf Acknowledgment}  \\
One of the authors (Y.S.) is thankful for the financial support 
 from Universit{\'e} Paris--Sud and also for the kind hospitality 
   during his stay at Ecole Normale  Sup\'erieure.  
This work was partially supported by a Grant-in-Aid for Scientific 
  Research from the Ministry of Education, Science, Sports and
  Culture (Grant No.09640429), Japan.

\appendix
%
%
\section{
Derivation of RG equations
}\label{app:ope_approach}
To derive RG equations, we use an Operator Product Expansion (OPE)
approach\cite{cardy_scaling}. in the absence of disorder, 
the relevant operator product expansions
are the following:
 \begin{eqnarray}\label{eq:ope_pure}
 \cos \sqrt{8} \phi
(x,\tau) \cos \sqrt{8} \phi(x',\tau') & \sim & \frac 1 2 \left[ 1- 4 \left[ (x-x')^2 (\partial_x
\phi)^2 \right. \right. \nonumber \\ 
 &+& \left. \left. (\tau-\tau')^2 (\partial_\tau \phi)^2 \right] \right] \nonumber \\
 \cos \sqrt{8} \theta
(x,\tau) \cos \sqrt{8} \theta(x',\tau') & \sim & 
  \frac 1 2 \left[ 1- 4 \left[ (x-x')^2 (\partial_x
\theta)^2\right. \right. \nonumber \\ 
& +& \left. \left. (\tau-\tau')^2 (\partial_\tau \theta)^2 \right]\right] 
\end{eqnarray}
To be able to make use of these OPEs, we must express
$\partial_{x,\tau} \theta$ as a function of $\partial_{x,\tau} \phi$. 
Using the equations on motion in imaginary time, we find:
\begin{eqnarray}
(\partial_x \theta)^2=-\frac{(\partial_\tau \phi)^2}{u^2 K^2}
\nonumber \\
(\partial_\tau \theta)^2=-\frac{u^2(\partial_x \phi)^2}{ K^2}
\end{eqnarray}
This leads to the following final form of the second OPE:
\begin{eqnarray}\label{eq:ope_pure_fin}
  \cos \sqrt{8} \theta
(x,\tau) \cos \sqrt{8} \theta(x',\tau') \sim \nonumber \\ 
\sim \frac 1 2 \left[ 1+ \frac{4}{K^2} \left[
\frac{(x-x')^2}{u^2}  (\partial_\tau
\phi)^2 + u^2(\tau-\tau')^2 (\partial_x \theta)^2 \right]\right]
\nonumber \\ 
\end{eqnarray}
Using Eq. (5:13) of Ref. \citen{cardy_scaling} 
we recover the RG equations of Refs.~\citen{giamarchi_spin_flop,nersesyan_2ch}. 

In the presence of disorder, we have to use replicas in order to
average over disorder. The replicated action (that we use only for the
purpose of perturbative RG) is the following:

\begin{eqnarray}
\lefteqn{}S_{\mathrm{random}} =  S_1+S_2 \nonumber \\
\lefteqn{}S_1=-\frac{D_s}{(\pi\alpha)^2}\int dx d\tau d\tau' \cos
\sqrt{2}\left[\phi^a_\rho(x,\tau)-\phi^b_\rho(x,\tau')\right]\times
\nonumber \\
\times  \cos
\sqrt{2}\phi^a_\parallel(x,\tau) \cos
\sqrt{2} \phi^b_\parallel(x,\tau') \nonumber \\
\lefteqn{}S_2 = -\frac{D_a}{(\pi\alpha)^2}\int dx d\tau d\tau' \cos
\sqrt{2}\left[\phi^a_\rho(x,\tau)-\phi^b_\rho(x,\tau')\right]\times
\nonumber \\ 
\times \cos
\sqrt{2}\theta^a_\parallel(x,\tau)  \cos
\sqrt{2}\theta^a_\parallel(x,\tau') \nonumber
\end{eqnarray}

The extra OPEs needed to derive the Renormalization Group equations in the
presence of disorder are:
\begin{eqnarray}
\cos \sqrt{2}\left[\phi^a_\rho(x,\tau)-\phi^b_\rho(x,\tau')\right] \cos
\sqrt{2}\phi^a_\parallel(x,\tau)\cos \sqrt{2}
\phi^b_\parallel(x,\tau') \sim \nonumber \\
\lefteqn{}  \delta_{ab}
\left\{ \frac 1 2
-\frac{(\tau-\tau')^2}{2} \left[(\partial_\tau \phi_\rho^a)^2 +
(\partial_\tau \phi_\parallel^a)^2\right] +\frac 1 2 \cos \sqrt{8}
\phi_a(x,\tau)\right\} \nonumber \\
\end{eqnarray}
\begin{eqnarray}
\cos \sqrt{2}\left[\phi^a_\rho(x,\tau)-\phi^b_\rho(x,\tau')\right] \cos
\sqrt{2}\theta^a_\parallel(x,\tau) \cos
\sqrt{2}\theta^b_\parallel(x,\tau') \sim \nonumber
\\ \lefteqn{} \delta_{ab}
\left\{ \frac 1 2
-\frac{(\tau-\tau')^2}{2} \left[(\partial_\tau \phi_\rho^a)^2 +
(\partial_\tau \phi_\parallel^a)^2\right] +\frac 1 2 \cos \sqrt{8}
\phi_a(x,\tau)\right\} \nonumber \\
\end{eqnarray}
\begin{eqnarray}
\cos \sqrt{2}\left[\phi_\rho^a(x,\tau)-\phi_\rho^b(x,\tau')\right] \cos
\sqrt{2} \phi^a_\parallel(x,\tau) \cos \sqrt{2}
\phi^b_\parallel(x,\tau')\nonumber \\
 \times \cos \sqrt{8}
\phi^b_\parallel(x'',\tau'') \sim \nonumber \\  \sim \frac 1 2 \cos \sqrt{2}\left[\phi_\rho^a(x,\tau)-\phi_\rho^b(x,\tau')\right] \cos
\sqrt{2} \phi^a_\parallel(x,\tau) \cos \sqrt{2}
\phi^b_\parallel(x,\tau') \nonumber \\
\end{eqnarray}
\begin{eqnarray} 
\cos \sqrt{2}\left[\phi_\rho^a(x,\tau)-\phi_\rho^b(x,\tau')\right] \cos
\sqrt{2} \theta_a(x,\tau) \cos \sqrt{2} \theta_b(x,\tau')\nonumber
 \\
\times \cos \sqrt{8}
\theta_b(x'',\tau'') \sim \nonumber \\
 \sim \frac 1 2 \cos \sqrt{2}\left[\phi_\rho^a(x,\tau)-\phi_\rho^b(x,\tau')\right] \cos
\sqrt{2} \theta^a_\parallel(x,\tau) \cos \sqrt{2}
\theta^b_\parallel(x,\tau') \nonumber \\ 
\end{eqnarray}

This leads to the RGE (\ref{eq:RGE_full}). 

\section{
Derivation of
localization-delocalization boundary  for the attractive interaction
}\label{app:criterion}

In the absence of disorder, the antisymmetric gap is given by the
coupled RG equations:

\begin{eqnarray}
\frac{d K_\parallel}{dl}=  \frac 1 2 (y_f^2 - K_\parallel^2 y_\perp^2 J_0(4K_\parallel\tilde{t}_\perp)) \nonumber \\
\frac{dy_f}{dl}=\left(2-\frac 2 {K_\parallel}\right) y_f   \nonumber \\
\frac{dy_\perp}{dl}=\left(2- 2 K_\parallel\right) y_\perp  \nonumber \\
\frac {d\tilde{t}_\perp}{dl}= \tilde{t}_\perp - \frac{y_\perp^2} 4
J_1(4K_\parallel\tilde{t}_\perp) 
\end{eqnarray}

To analyze these equations, we remark that if $\tilde{t}_\perp=0$,
$y_\perp=y_f$ and $K_\parallel=1$ there is no flow in the RG
equations. Let us consider the limit $\tilde{t}_\perp(0) \ll 1$. 
In such case, we can make the approximation:
$J_0(4K_\parallel\tilde{t}_\perp)\simeq 1$ for small $l$. As a result, 
there is no flow of $K_\parallel,y_f,y_\perp$. Nevertheless,
$\tilde{t}_\perp$ flows as $\tilde{t_\perp}(l)=\tilde{t}_\perp(0)
\e^l$ (we have neglected $y_\perp^2$ compared to $1$). 
The approximation $J_0(4K_\parallel\tilde{t}_\perp)\simeq 1$ will
therefore break down at the scale $l^*$ when $4K_\parallel\tilde{t}_\perp(0)
\e^{l^*}\simeq 1$. 
Beyond the scale $l^*$, we make the assumption that
$J_0(4K_\parallel\tilde{t}_\perp)\simeq 0$. Then, $y_\perp$ disappears 
from the RG equation of $K_\parallel$ and the only remaining RG
equations are:
\begin{eqnarray}\label{eq:KT_equation}
\frac{d K_\parallel}{dl}=  \frac  {y_f^2} 2  \nonumber \\
\frac{dy_f}{dl}=\left(2-\frac 2 {K_\parallel}\right) y_f   \nonumber \\
\end{eqnarray}

The initial conditions for these RG equations are with our
approximations: 
\begin{eqnarray}
K_\parallel(l^*)=K_\parallel(0)=1 \virg \nonumber \\
y_f(l^*)=y_f(0)\;\; .
\end{eqnarray}

It is convenient to introduce the variable $y_\parallel$ in
(\ref{eq:KT_equation}) defined by: $K_\parallel=1+\frac{y_\parallel} 2$ so 
that the RG equations can be rewritten:
\begin{eqnarray}\label{eq:KT_canonical}
\frac{dy_\parallel}{dl}=y_f^2 \nonumber \\
\frac{dy_f}{dl}=y_\parallel y_f 
\end{eqnarray}

With initial conditions: $y_\parallel(l^*)=0$.
 A convenient parameterization of the RG equations (\ref{eq:KT_canonical}) is~:
\begin{eqnarray}
y_\parallel(l)=y_f(0) \sinh \theta(l) \nonumber \\
y_f(l)=y_f(0) \cosh \theta(l)
\end{eqnarray}
with initial condition $\theta(l^*)=0$. The system
(\ref{eq:KT_canonical}) is the reduced to a single differential
equation:
\begin{equation}
\frac {d\theta}{dl}=y_f(0) \cosh \theta
\end{equation} 
which leads to:
\begin{equation}
2(\arctan(\e^{\theta(l)})-\frac \pi 4)=y_f(0)(l-l^*)
\end{equation}
The RG flow is cut when $y_f(\tilde{l})\simeq 1$, i.e. for $y_f(0)
\e^{\theta(\tilde{l})}/2\simeq 1$. 
This leads to $\tilde{l}-l^*=\frac{\pi}{2 y_f(0)}-1$. 
The correlation length $\xi_\parallel$ is then $a \e^{\tilde{l}}=a
\e^{\tilde{l}-l^*}\e^{l^*}$, i.e. 
\begin{equation}
\xi_\parallel= a \exp \left(\frac{\pi}{2 y_f(0)}\right) \frac{1}{4\e
\tilde{t}_\perp(0)} 
\end{equation}
The resulting gap is then $\Delta_\parallel=\frac {u_\parallel}{\xi}$
thus:
\begin{equation}
\Delta_\parallel=4 \e \tilde{t}_\perp \exp \left(-\frac{\pi}{2 y_f(0)}\right)
\end{equation}

As long as we can neglect disorder we will obtain the expression of
$\Delta_\parallel$ of the preceding section. in order to be able to
neglect disorder, we need $\xi_{{\rm 1 ch.}} \gg \xi_\parallel$. This
implies the criterion:

 \begin{equation}\label{eq:critical_criterion}
{\cal D}^{1/(3-2K)} \ll 4 \e \tilde{t}_\perp \exp \left(-\frac{\pi}{2
y_f(0)}\right) 
\end{equation} 

Thus, we expect the critical disorder ${\cal D}_c$ to be given by:
\begin{equation}\label{eq:critical_disorder}
{\cal D}_c^{1/(3-2K)} = C_0 4 \e \tilde{t}_\perp \exp \left(-\frac{\pi}{2
y_f(0)}\right) 
\end{equation} 
where $C_0$ is a numerical prefactor.


\end{document}